\documentclass[orivec, envcountsame, runningheads]{llncs}

\usepackage{amsmath}
\usepackage{amssymb}
\usepackage{stmaryrd} 
  \SetSymbolFont{stmry}{bold}{U}{stmry}{m}{n} 
\usepackage{xspace} 
\usepackage{cite}
\usepackage{rotating}
\usepackage{combelow} 
\usepackage{paralist}  

\usepackage[kerning=true]{microtype} 

\usepackage{subfigure}
\usepackage{tikz}
  \usetikzlibrary{calc,shapes,arrows}
\usepackage{xhfill}
\usepackage{textcomp} 

\newcommand{\ignore}[1]{}

\renewcommand{\phi}{\varphi}
\renewcommand{\rho}{\varrho}

\newcommand{\TRUE}{\mathtt{TRUE}}

\newcommand{\NOT}{\mathtt{NOT}}

\newcommand{\OR}{\mathtt{OR}}
\newcommand{\IMPLIES}{\mathtt{IMPLIES}}
\newcommand{\PREVIOUS}{\mathtt{PREVIOUS}}
\newcommand{\NEXT}{\mathtt{NEXT}}
\newcommand{\SINCE}{\mathtt{SINCE}}

\newcommand{\UNTIL}{\mathtt{UNTIL}}
\newcommand{\EVENTUALLY}{\mathtt{EVENTUALLY}}
\newcommand{\ALWAYS}{\mathtt{ALWAYS}}
\newcommand{\FREEZE}{\mathtt{FREEZE}}

\newcommand{\true}{\mathsf{t}}
\newcommand{\false}{\mathsf{f}}
\newcommand{\unknown}{\bot}

\DeclareFontFamily{U}{mathb}{\hyphenchar\font45}
\DeclareFontShape{U}{mathb}{m}{n}{
      <5> <6> <7> <8> <9> <10> gen * mathb
      <10.95> mathb10 <12> <14.4> <17.28> <20.74> <24.88> mathb12
      }{}
\DeclareSymbolFont{mathb}{U}{mathb}{m}{n}

\DeclareMathSymbol{\sqsubset}{3}{mathb}{"80}
\DeclareMathSymbol{\sqsubseteq}{3}{mathb}{"84}
\DeclareMathSymbol{\sqsubsetneq}{3}{mathb}{"88}

\newcommand{\POLIMON}{\mbox{POL{\'I}MON}}
\newcommand{\MTLfreeze}{\mbox{MTL$^\downarrow$}}

\title{\POLIMON: Checking Temporal Properties over Out-of-order Streams at Runtime}
\titlerunning{\POLIMON}

\author{Felix Klaedtke}
\authorrunning{Klaedtke}
\institute{NEC Laboratories Europe, Heidelberg}

\allowdisplaybreaks

\begin{document}

\maketitle

\begin{abstract}
  This paper presents the monitoring tool \POLIMON{} for checking
  system behavior at runtime against specifications expressed as
  formulas in the real-time logic MTL or its extension with the freeze
  quantifier.  The tool's distinguishing feature is that \POLIMON{}
  can receive messages describing the system events out of order.
  Furthermore, since \POLIMON{} processes received messages
  immediately, it outputs verdicts promptly when a message's described
  system event leads to a violation of the specification.  This makes
  the tool well suited, e.g., for verifying the behavior of
  distributed systems with unreliable channels at runtime.
\end{abstract}

\setlength{\abovedisplayskip}{4pt plus 3pt minus 4pt}
\setlength{\belowdisplayskip}{4pt plus 3pt minus 4pt}

\section{Introduction}
\label{sec:intro}

Monitoring system behavior and checking it during runtime is widely
used for a diversity of systems and properties.  An important class of
properties are requirements on the temporal occurrence and
nonoccurrence of system events.  For instance, requests must be served
within a given time period and a failed login must not be directly
followed by another login attempt.  Linear-time temporal logics like
LTL or variants and extensions thereof are well suited to formally
express such requirements and various online algorithms have been
developed for them for the verification of system behavior at runtime,
see,
e.g.,~\cite{Havelund_Rosu:monitoring_safty,Bauer_etal:rv_tltl,Rosu_Chen:slicing,Basin_etal:rv_mfotl}.

A major challenge in checking system behavior at runtime is when the
system under scrutiny is composed of several interacting, distributed
components~\cite{Basin_etal:outoforder_journal,ColomboFalcone:global_clock,Sen_etal:decentralized_disitributed_monitoring,Scheffel_Schmitz:asynchronous_distributed_rv}.
Most of today's IT systems are of this form.  For instance, the
observations of a monitor may only be partial or received out-of-order
because of network latencies or system component failures.  Note that
related challenges arise in stream processing applications where data
must be queried or aggregated in real time, is not necessarily
complete, and may be received in an order different from is
generation~\cite{Stonebraker_etal:requirements}.

\vspace{-.2cm}
\paragraph{Contributions.}

In this paper, we present the tool \POLIMON{} for monitoring
out-of-order event streams.  Concretely, the tool's objective is to
determine at runtime whether system behavior, as observed and reported
by system components, satisfies or violates a given specification.
\POLIMON{} handles specifications expressed as formulas in the
real-time logic MTL and its extension with the freeze quantifier
\MTLfreeze{} (pronounced ``MTL freeze''). The freeze quantifier allows
one to bind logical variables to data values occurring in an event.  A
cornerstone of \POLIMON's underlying monitoring
approach~\cite{Basin_etal:outoforder_journal} is a three-valued
semantics for MTL and \MTLfreeze{} that is well suited to reason in
settings where system components communicate with the monitors over
unreliable channels.  \POLIMON's reported verdicts are sound and
complete with respect to the tool's partial knowledge about the system
behavior.

\POLIMON{} operates as follows.  It receives as input timestamped
messages from the system components that describe the system events.
No assumptions are made on the order in which these messages are
received.
For each received message, \POLIMON{} updates its state.  This state
comprises an acyclic graph structure for reasoning about the system
behavior, that is, computing verdicts about the monitored
specification's fulfillment.  The graph's nodes store the truth values
of the subformulas for the different times with data values frozen to
logical variables, including the times with no or only partial
knowledge.  \POLIMON{} refines the graph when receiving information
about a specific point in time in form of events.
In each such update, data values are propagated down to the graph's
leaves and Boolean truth values for subformulas are propagated up
along the graph's edges.  When a Boolean truth value is propagated to
a root node of the graph, \POLIMON{} outputs the corresponding
verdict.

\POLIMON{} is significantly faster than its
predecessor~\cite{Basin_etal:outoforder_journal}.  The speedup is
rooted in various algorithmic improvements on maintaining the graph
structure.  Furthermore, \POLIMON{} utilizes multiple CPU cores by
processing the received messages in a pipeline.  \POLIMON{} also
supports new features.  For instance, \POLIMON{} can export and import
its internal state, which are crucial operations for
migrating monitors.  \POLIMON{} also implements different approaches
for closing knowledge gaps.

\vspace{-.2cm}
\paragraph{Related Tools.}

Various monitoring tools have been developed within the
runtime-verification community over the last years, for instance,
EAGLE~\cite{Barringer_etal:eagle}, MOP~\cite{Meredith_etal:mop},
BeepBeep~\cite{beepbeep}, MONPOLY~\cite{monpoly}, AMT~\cite{amt,amt2},
{\sc Montre}~\cite{montre}, and StreamLAB~\cite{streamlab}, each with
unique features.  Many of these tools handle---as
\POLIMON---spe\-ci\-fi\-ca\-tions expressed in some formalism that extends the
linear-time temporal logic LTL.
Metric constraints are, e.g., supported by the tools MONPOLY, AMT,
{\sc Montre}, and \POLIMON.
EAGLE and \POLIMON{} both support the freeze quantifier to reason
about the events' data values.  Freeze quantification in turn is a
weak form of first-order quantification, which is supported by
BeepBeep and MONPOLY.  Note that the MOP framework also supports some
reasoning about data by slicing the traces according to the events'
data values.
There are, however, often subtle differences in the tools' underlying
time models and the semantics of their specification languages.
Concretely, the semantics of \POLIMON's specification language uses a
three-valued semantics, which conservatively approximates the
classical Boolean semantics.  \POLIMON's unique feature is that it
soundly and completely deals with knowledge gaps.

\vspace{-.2cm}
\paragraph{Organization.}

The remainder of this paper is structured as follows.  In
Section~\ref{sec:tool}, we provide an overview of the tool.  In
Section~\ref{sec:eval}, we report on the tool's performance.  Finally,
in Section~\ref{sec:concl}, we draw conclusions.

\section{\POLIMON{} in a Nutshell}
\label{sec:tool}

\begin{figure}[t]
  \centering
  \vspace{-.4cm}
  \includegraphics[scale=.11]{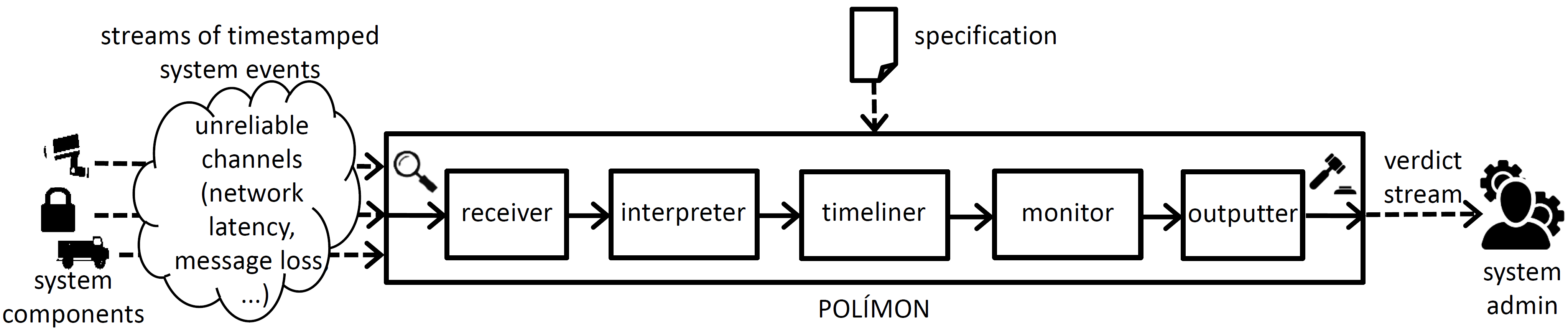}
  \vspace{-.4cm}
  \caption{Tool overview.}
  \label{fig:overview}
  \vspace{-.4cm}
\end{figure}

\POLIMON's input comprises (1)~a specification and (2)~event streams
from the monitored system components.  See also
Figure~\ref{fig:overview}.  \POLIMON{} processes the incoming events,
which it either receives over a UDP socket or reads from a file,
iteratively and checks them against the specification, which is a
formula in the real-time logic
MTL~\cite{Koymans:realtime_properties,AlurHenzinger:realtimelogics_survey}
extended with the freeze quantifier.
\POLIMON{} does not require that the events are received in the order
in which they are generated.  Each event is processed immediately,
that is, \POLIMON{} interprets the incoming message that describes the
event, updates the monitor state, and outputs the computed verdicts.

We proceed by first illustrating \POLIMON's usage through a simple
example before we describe the tool's theoretical underpinnings and
its main components, including some details on their implementation.

\subsection{Example}

Consider a system in which agents can issue tickets. Issuing a ticket
is represented by an event $\mathit{ticket}(a,n)$, where $a$ is the
agent who issued the ticket and $n$ is the ticket
identifier. Furthermore, consider the simple specification that an
agent must wait for some time (e.g. 100 milliseconds) before issuing
another ticket.  A formalization in \POLIMON's specification language
is as follows.
\begin{equation*}
  \begin{array}{@{}l@{}l@{}}
    \FREEZE\ a[agent], n[\mathit{id}].\
    &\mathit{ticket}(a,n)
    \\
    &\IMPLIES
    \\
    &\NOT\ \EVENTUALLY(0,100\mathit{ms}]\
    \FREEZE\ m[\mathit{id}].\ \mathit{ticket}(a,m)
  \end{array}
\end{equation*}
The $\FREEZE$ quantifiers bind the logical variables to the data
values that occur in the events.  The data values are stored in
predefined registers ($\mathit{agent}$ and $\mathit{id}$ in the above
formula) and each event determines the register values at the event's
occurrence.  Note that the data values from an event bound by the
outer $\FREEZE$ quantifier in the above formula stem from an event
that is different from the one for the inner $\FREEZE$ quantifier,
since the metric constraint of the temporal future-time connective
$\EVENTUALLY$ excludes $0$.  Furthermore, note that an event also
determines the interpretation of the predicate symbol
$\mathit{ticket}$ at the event's occurrence, namely, the singleton
$\{(a,n)\}$.
Finally, note that there is an implicit outermost temporal connective
$\ALWAYS$.  For each received event, a reported verdict describes
whether the given specification is violated or satisfied at the
event's time.

\POLIMON{} requires that events are timestamped (in Unix time with a
precision up to microseconds).  \POLIMON{} additionally requires that
each event includes the event's system component and a \emph{sequence
  number}, i.e., the $i$th event of the component.  \POLIMON{} uses
the sequence numbers to determine whether it has received all events
from a component within a given time period.  For instance, when
\POLIMON{} is monitoring a single component and it has received events
with the sequence number $i$ and $i+2$ but no event with the sequence
number $i+1$, then \POLIMON{} knows that there is a \emph{knowledge
  gap} between the two events with the sequence numbers $i$ and $i+2$.

For our example, assume that \POLIMON{} is monitoring the single
component~$\mathtt{syscomp}$ and receiving the following events from
it.
\begin{equation*}
  \begin{array}{@{}l@{}}
    1548694551.904@[\mathtt{syscomp}]\ (1): \mathit{ticket}(\mathtt{alice},34)
    \\
    1548694551.996@[\mathtt{syscomp}]\ (2): \mathit{ticket}(\mathtt{bob},8)
    \\
    1548694552.059@[\mathtt{syscomp}]\ (3): \mathit{ticket}(\mathtt{charlie},52)
    \\
    1548694552.084@[\mathtt{syscomp}]\ (5): \mathit{ticket}(\mathtt{bob},11)
    \\
    1548694552.407@[\mathtt{syscomp}]\ (6): \mathit{ticket}(\mathtt{charlie},1)
    \\
    1548694552.071@[\mathtt{syscomp}]\ (4): \mathit{ticket}(\mathtt{charlie},99)
  \end{array}
\end{equation*}
Note that the events are received in the order of their timestamps,
except the event with the sequence number~$4$ is received out of
order.  The delay of receiving this event might, e.g., be caused by
network latencies. \POLIMON{} processes the events in the order it
receives them.

\POLIMON{} does not output any verdict after receiving the first two
events.  When processing the third event, \POLIMON{} outputs the
verdict
\begin{equation*}
  \mathit{true}@1548694551.904
  \,.
\end{equation*}
The reason is that enough time has elapsed for $\mathtt{alice}$ to
issue another ticket.  This inference is sound since no events are
missing up to the third event and the time difference between the
first event and the third event is larger than 100~milliseconds.
The next verdict \POLIMON{} outputs is when processing the second
event with agent $\mathtt{bob}$:
\begin{equation*}
   \mathit{false}@1548694551.996
  \,.
\end{equation*}
Note that although there is a knowledge gap (the event with the
sequence number~$4$ is missing), outputting this verdict is sound,
since no matter how this gap is filled, the second event with
agent~$\mathtt{bob}$ causes a violation of the specification for the
first event with agent~$\mathtt{bob}$.
When receiving the second last event, \POLIMON{} outputs the verdict
\begin{equation*}
  \mathit{true}@1548694552.084
\end{equation*}
but not the verdict $\mathit{true}@1548694552.059$.  Outputting the
latter verdict would not be sound because of the knowledge gap between
the events with sequence numbers~$3$ and~$5$.  In fact, the event,
which arrives late, causes a violation.  Finally, when receiving this
last event, \POLIMON{} outputs the following two verdicts.
\begin{equation*}
  \begin{array}{@{}l@{}}
    \mathit{true}@1548694552.071
    \\
    \mathit{false}@1548694552.059
  \end{array}
\end{equation*}

We have used \POLIMON{} in checking the interactions of components in
software-defined networks (SDN controllers, network applications, and
switches).  In particular, \POLIMON{} receives events describing the
actions performed by the various network components, and checks in
real time, e.g., whether certain deadlines are met.  In case of
violations, components can be \mbox{reconfigured or isolated.}

\subsection{Foundations}

\POLIMON's theoretical underpinnings, which we summarize in the
following, are described in detail
in~\cite{Basin_etal:outoforder_journal}.

Specifications are given as formulas in the real-time logics MTL or
\MTLfreeze.  \POLIMON's core syntax is given by the grammar
\begin{equation*}
  \begin{array}{@{}lcl@{}}
    \varphi &\ \mathbin{::=}\ &
    \TRUE \ \big|\
    p(x_1,\dots,x_n) \ \big|\
    \NOT\ \phi \ \big|\
    \phi \ \OR\ \phi \ \big|\
    \FREEZE\ x_1[r_1],\dots,x_n[r_n].\ \phi \ \big|\
    \\[.1cm]
    & &
    \PREVIOUS\,I\ \phi \ \big|\
    \NEXT\,I\ \phi \ \big|\
    \phi\ \SINCE\,I\ \phi \ \big|\
    \phi\ \UNTIL\,I\ \phi
    \,,
  \end{array}
\end{equation*}
where the $x_i$s range over variable names, $p$ ranges over the
predicate symbols, the $r_i$s over register names, and $I$ ranges over
intervals, e.g., $[0,*)$ is the unbounded left-closed interval with
left bound $0$.  \POLIMON{} supports various standard syntactic sugar,
which it unfolds.  For instance, $\phi\ \IMPLIES\ \psi$ abbreviates
$(\NOT\ \phi)\ \OR\ \psi$ and $\EVENTUALLY\ \phi$ abbreviates
$\TRUE\ \UNTIL[0,*)\ \phi$.  \POLIMON{} also supports rigid predicates
like the ordering on integers, which we omit here for the sake of
brevity.

The semantics of the specification language follows intuitively the
classical point-based semantics for
MTL~\cite{Koymans:realtime_properties,AlurHenzinger:realtimelogics_survey}.
However, for dealing with knowledge gaps, the semantics is not defined
over timed words but over so-called
\emph{observations}~\cite{Basin_etal:outoforder_journal}. Observations
are finite words in which knowledge gaps are made explicit.
Furthermore, the semantics is defined over three truth values $\true$,
$\false$, and $\unknown$, which are interpreted as in strong Kleene
logic, namely, they are partially ordered ($\unknown$ is less than
$\true$ and $\false$, which are incomparable) and $\true$ stands for
true, $\false$ for false, and $\unknown$ for unknown.  The semantics
conservatively approximates the classical Boolean semantics.  We refer
to~\cite{Basin_etal:outoforder_journal} for details.

The reason of using a nonstandard semantics is for providing both
soundness and completeness guarantees of the verdicts in the presence
of knowledge gaps.  Intuitively speaking, \emph{soundness} means that
\POLIMON{} only outputs a verdict when this verdict remains valid no
matter how gaps are filled later, and \emph{completeness} means that
\POLIMON{} outputs a verdict as soon as possible based on the already
received events.  We note that there is always at least one knowledge
gap, even when events are received inorder, namely, a \POLIMON{} does
not know what events occur after the event with the largest timestamp.

\POLIMON{} makes the following system assumptions.  First, a system's
behavior is described completely through infinitely many events. Note
that \POLIMON{} does not require to receive all of them in the
limit. Second, the monitored system components are fixed and known to
\POLIMON.  Third, components do not send bogus events.
\POLIMON's underlying time model is based on wall-clock time.
\POLIMON{} requires that the events' timestamps are unique and the
events are linearly ordered by their timestamps.  Note that this is a
strong requirement, since in particular for distributed systems,
events can actually only be partially ordered by logical clocks.  If
events are, however, only partially ordered, the reasoning becomes
significantly more difficult since a finite partially ordered trace
can have exponentially many interleavings.  Furthermore, we argue that
when events are generated at the speed of microseconds, existing
protocols like the Network Time Protocol~(NTP) work well in practice
to synchronize clocks between distributed components.  This
justification becomes however questionable when events are generated
at a very high speed, e.g., in the low microsecond range or even in
the nanosecond range.

\subsection{Implementation Details}
\label{subsec:impl_details}

\POLIMON{} is a command-line tool and written in the programming
language Go (\url{www.golang.org}).\footnote{\POLIMON{} has not been
  opensourced.  However, for nonprofit use, it can be obtained freely
  under a source code license agreement from NEC Laboratories Europe
  GmbH.}  In the following, we provide an overview of its
implementation.

\POLIMON{} processes incoming events in a pipeline, where the pipeline
stages are executed as separate goroutines.  See
Figure~\ref{fig:overview}.
The first stage, the \emph{receiver}, parses the incoming events.
The second stage, the \emph{interpreter}, extracts the
event's data values, and determines the interpretation of the
predicate symbols at the events' time point via regular-expression
matching.
The third stage, the \emph{timeliner}, determines based on the
events' sequence numbers the time periods for which all events have
been received.
The fourth stage, the \emph{monitor}, computes the verdicts.
Finally, the fifth stage, the \emph{outputter}, reports the verdicts.

\POLIMON{} supports different implementations of these stages, e.g.,
receivers and outputters for different input and output formats.
\POLIMON{} currently also comprises two monitors: one for MTL and one
for \MTLfreeze.  Furthermore, \POLIMON{} implements multiple
timeliners: one in which events must be received inorder and two for
receiving events out of order.  The first one uses sequence numbers to
determine whether there are gaps between messages and the second one
uses \emph{watermark} messages~\cite{flink} to close gaps between
messages.  Watermark messages are control messages that are inserted
into the pipeline.  Another type of control messages that can be
inserted into the pipeline are messages for outputting the internal
states of the timeliner and the monitor.  The outputter is responsible
for writing the state as a JSON object into a file.

\POLIMON's key data structure is an acyclic graph that represents the
intermediate specification's fulfillment with respect to the partial
knowledge about the system behavior.  This graph can be understood as
a combinational circuit, which is refined for each received event.
Such a refinement comprises the steps of (1)~splitting gates, (2)~the
downward propagation of data values, which may create new gates, and
(3)~the upward propagation of Boolean truth values.
The gates are arranged along three axes: The x-axis is the temporal
dimension, the y-axis the formula structure, and the z-axis the data
dimension (which is trivial for propositional formulas).  Each
variable assignment corresponds to a plane along the z-axis.

\begin{figure}[t]
    \centering
    \vspace{-.4cm}
    \scalebox{.7}{
\begin{tikzpicture}[thick, x=1pt, y=1pt]
  \tikzstyle{node} = [draw, minimum height = 14pt, minimum width = 14pt]
  \tikzstyle{trigger} = [->]
  \tikzstyle{implicittrigger} = [dashed,->]
  \tikzstyle{guard} = [draw, circle, fill, minimum size = 0pt, inner sep = 0pt]

  \newcommand{\x}{0}
  \node at (\x,40) {$\EVENTUALLY(0,3]\ p$};
  \node at (\x,0) {$\phantom{\EVENTUALLY(0,3]\ }p$};
  
  \renewcommand{\x}{80}
  \node at (\x,-25) {(a) initial};
  
  \draw (\x-20,60) -- node[above] {$[0,\infty)$} ++ (40,0);
  
  \node[node] (Neventually) at (\x,40) {$ $};
  \node[node] (Natomic) at (\x,0) {$\ $};
  
  \node[guard] (Geventually) at (\x,33) {};
  \draw[trigger] (Natomic) to [out=90, in=-90] (Geventually);        
  
  \renewcommand{\x}{160}
  \node at (\x+50,-25) {(b) split for an event with timestamp $\tau$};
  
  \draw (\x-10,60) -- node[above] {$[0,\tau)$} ++ (40,0);
  \draw[|-|] (\x+50,60) -- node[above] {$\phantom{\{}\tau\phantom{\}}$} ++ (0.1,0);
  \draw (\x+70,60) -- node[above] {$(\tau,\infty)$} ++ (40,0);
  
  \node[node] (NeventuallyL) at (\x+10, 40) {$\ $};
  \node[node] (NatomicL) at (\x+10, 0) {$\ $}; 
  
  \node[node] (NeventuallyM) at (\x+50, 40) {$\ $};
  \node[node] (NatomicM) at (\x+50, 0) {$\ $};
  
  \node[node] (NeventuallyR) at (\x+90, 40) {$\ $};
  \node[node, fill=white] (NatomicR) at (\x+90, 0) {$\ $};
  \node[guard] (GatomicR) at (\x+96, 7) {};
  
  \node[guard] (GeventuallyL1) at (\x+4, 33) {};
  \node[guard] (GeventuallyL2) at (\x+10, 33) {};
  \node[guard] (GeventuallyL3) at (\x+16, 33) {};
  
  \node[guard] (GeventuallyM) at (\x+61, 42) {};
  
  \node[guard] (GeventuallyR) at (\x+90, 33) {};
  
  \draw[trigger] (NatomicL) to [out=90, in=-90] (GeventuallyL1); 
  \draw[trigger] (NatomicM) .. controls (\x+50,20)  and (\x+10,0) .. (GeventuallyL2);
  \draw[trigger] (GatomicR) to [out=90, in=-90] (NeventuallyR.south);
  \draw[implicittrigger] (NatomicR) to [out=90, in=-90] (NeventuallyM.south);
  \draw[implicittrigger] (NatomicR) .. controls (\x+70,30) and (\x+16,5) .. (GeventuallyL3);
  
  \draw[dashed] (NeventuallyL) to [out=0, in=180] (NeventuallyM);
  \draw[dashed] (NeventuallyM) to [out=0, in=180] (NeventuallyR);
  
  \draw[dashed] (NatomicL) to [out=0, in=180] (NatomicM);
  \draw[dashed] (NatomicM) to [out=0, in=180] (NatomicR);
\end{tikzpicture}
    }
    \vspace{-.4cm}
    \caption{Graph-based monitoring data structure.}
    \label{fig:datastruct}
    \vspace{-.4cm}
\end{figure}
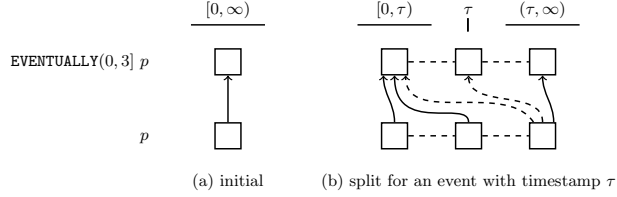

Figure~\ref{fig:datastruct} illustrates this graph-based data
structure for $\EVENTUALLY(0,3]\ p$ by showing (a)~the initial graph
that comprises a node for each subformula for the interval
$[0,\infty)$, and (b)~the split when receiving an event with the
timestamp~$\tau$; Boolean values have not been propagated.
Note that in Figure~\ref{fig:datastruct}(b) the initial knowledge
gap~$[0,\infty)$ is split into a left and a right knowledge gap,
$[0,\tau)$ and $(\tau,\infty)$, respectively. Furthermore, a time
point for the received event is added. The new intervals and the time
point inherit the nodes of the ancestor interval and the nodes' edges
are adjusted according to the semantics of the temporal connective and
its metric constraint.  The dashed edges are implicit. They are
derived by following the solid edges and the nodes' horizontal
successors or predecessors.  This small detail is actually critical
for \POLIMON's performance (see Section~\ref{sec:eval}): each node has
a single outgoing edge and splits are therefore more local.
After a split, Boolean truth values are propagated up along the edges.
There are several subtle cases for the propagation.  For instance, in
Figure \ref{fig:datastruct}(b), the propagation of a Boolean truth
value of the proposition $p$ at time $\tau$ can only be carried out
when the metric constraint is valid, i.e., if $\tau\leq3$.

\section{Experimental Evaluation}
\label{sec:eval}

In this section, we report on \POLIMON's performance.  For the
experiments, we used a computer with an Intel Xeon 3.3\,GHz CPU (four
physical cores).  \POLIMON{} was compiled with the Go compiler
version~1.13.x.

\begin{figure}[t]
  \centering
  \vspace{-.4cm}
  \includegraphics[scale=.47]{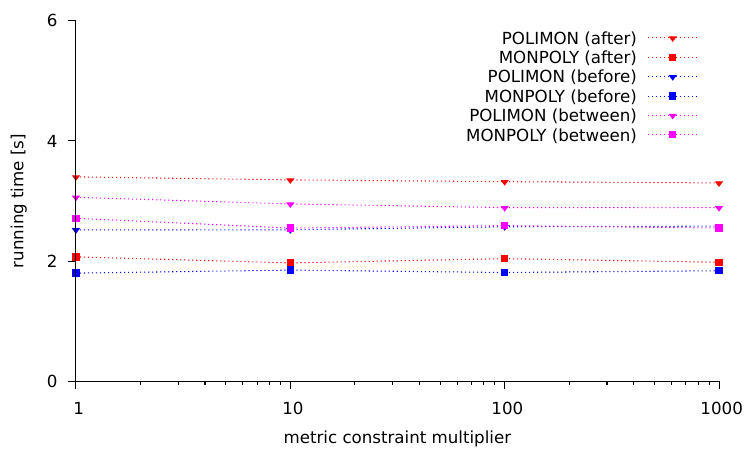}
  \
  \includegraphics[scale=.47]{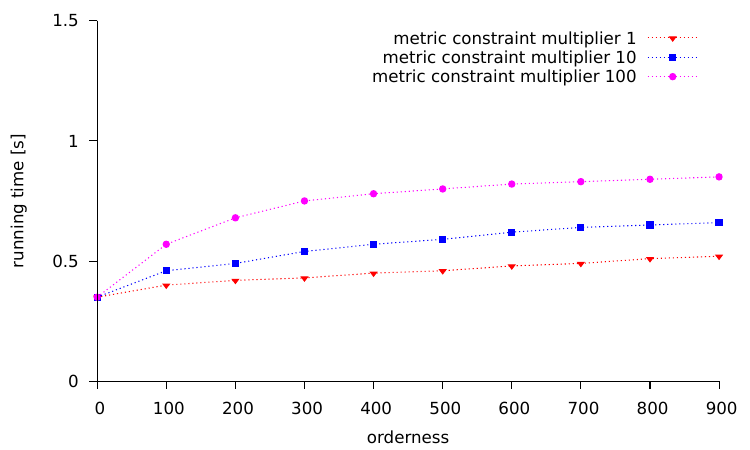}
  \vspace{-.6cm}
  \caption{Running times for the Timescales pattern family {\sf
      past-always} comprising three formulas on ordered event streams
    with $10^6$ time points (left) and running times for the pattern
    {\sf past-always-after} on out-of-order event streams with $10^5$
    time points (right).}
  \label{fig:timescales}
  \vspace{-.2cm}
\end{figure}

In a first set of experiments, we used the Timescale benchmark
suite~\cite{timescales}, which includes log generators for 20 temporal
parameterized propositional patterns with metric constraints.  We also
evaluated \POLIMON's performance on out-of-order event streams that we
obtained from the ordered logs generated by Timescales by adding
artificial ``arrival delays'' to each event's timestamp.
The plots in Figure~\ref{fig:timescales} show \POLIMON's running times
for a few Timescales patterns on ordered and out-of-order event
streams with different metric constraints.  For comparison, the plot
on the left-hand side of Figure~\ref{fig:timescales} includes the
running times of the MONPOLY tool.  Our results for the other
Timescales patterns are similar.
Overall, \POLIMON's performance is competitive to MONPOLY's
performance on ordered logs.  Note that some patterns cannot be
handled by MONPOLY because of the tool imposes the syntactic
restriction that temporal future-time connectives must not be
unbounded.  Handling out-of-order event streams does not significantly
increase running times.  The reason is that most updates on the graph
structure are local and thus involve only a few nodes. Recall from
Section~\ref{subsec:impl_details} that each node has only a single
outgoing edge, which needs to be updated; the implicit edges in
contrast often follow from the single outgoing edge without further
computations.  We also measured the tools' memory usage.  In general,
memory usage is low for both tools.  \POLIMON's memory usage usually
increases slowly with the ``out-of-orderness'' in event streams.

In a second set of experiments, we used MTL and \MTLfreeze{} formulas
expressing policies for handling suspicious financial transactions.
The formulas are similar to the ones used in previous evaluations of
monitoring
tools~\cite{Basin_etal:rv_mfotl,Basin_etal:outoforder_journal}.
We again generated event streams synthetically for different event rates
(i.e., events per second) and, similar as in the Timescales
experiments, we obtained out-of-order streams by adding artificial
arrival delays to each event's timestamp taken.  The delays were taken
from a normal distribution for different standard deviations.
\begin{figure}[t]
  \centering
  \vspace{-.4cm}
  \scalebox{.75}{\begin{minipage}{16cm}
  \begin{gather}
    \tag{Q1}
    \mathit{suspicious} \ \IMPLIES\ \EVENTUALLY[0,3]\ \mathit{report}
    \\
    \tag{Q2}
    \mathit{suspicious} \ \IMPLIES\ \ALWAYS(0,5]\ \NOT\ \mathit{suspicious}
    \\
    \tag{Q3}
    \mathit{suspicious} \ \IMPLIES\
    \big((\mathit{transaction}\ \IMPLIES\ \EVENTUALLY[0,3]\ \mathit{report}) \ \UNTIL\ \mathit{unflag} \big)
    \\
    \tag{Q4}
    \mathit{suspicious} \ \IMPLIES\
    \ALWAYS[0,6]\ (\mathit{transaction} \ \IMPLIES\ \EVENTUALLY[0,3]\ \mathit{report})
  \end{gather}
  \vspace{-.7cm}
  \par\xhrulefill{black}{0.1pt}\par
  \vspace{-.5cm}
  \begin{gather}
    \tag{P1}
    \mathit{suspicious}\ \IMPLIES\ \FREEZE\ t[\mathit{tid}].\ \EVENTUALLY[0,3]\ \mathit{report}(t)
    \\
    \tag{P2}
    \mathit{suspicious}\ \IMPLIES\
    \FREEZE\ c[\mathit{cid}].\ \ALWAYS(0,5]\ \big(\mathit{suspicious}\ \IMPLIES\ \NOT\ \mathit{transaction}(c)\big)
    \\
    \tag{P3}
    \mathit{suspicious}\ \IMPLIES\
    \FREEZE\ c[\mathit{cid}],\, t[\mathit{tid}].\
    \big((\mathit{transaction}(c)\ \IMPLIES\ \FREEZE\ s[\mathit{tid}].\ s=t) \ \UNTIL\  \mathit{unflag}(c)\big)
    \\
    \tag{P4}
    \mathit{suspicious}\ \IMPLIES\
    \FREEZE\ c[\mathit{cid}].\
    \ALWAYS[0,6]\ \big(\mathit{transaction}(c)\ \IMPLIES\ \FREEZE\ t[\mathit{tid}].\ \EVENTUALLY[0,3]\ \mathit{report}(t)\big)
  \end{gather}
  \end{minipage}}
  \vspace{-.2cm}
  \caption{Formulas used in the experimental evaluation.}
  \label{fig:formulas}
  \vspace{-.2cm}
\end{figure}

\begin{figure}[t!]
  \centering
  \vspace{-.4cm}
  \subfigure[inorder (MTL)]{\includegraphics[scale=.47]{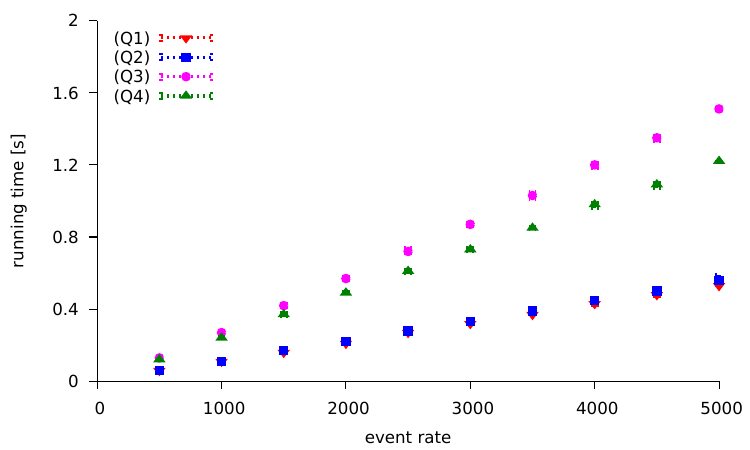}}
  \
  \subfigure[out of order (MTL, event rate~$1000$)]{\includegraphics[scale=.47]{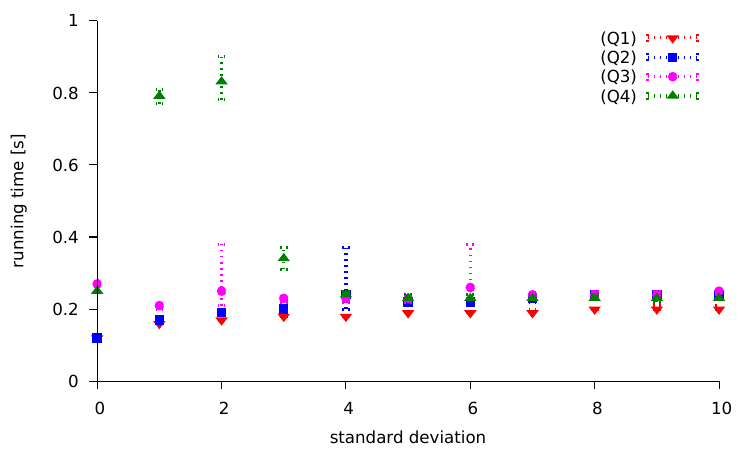}}
  \\[-.2cm]
  \subfigure[inorder (\MTLfreeze)]{\includegraphics[scale=.47]{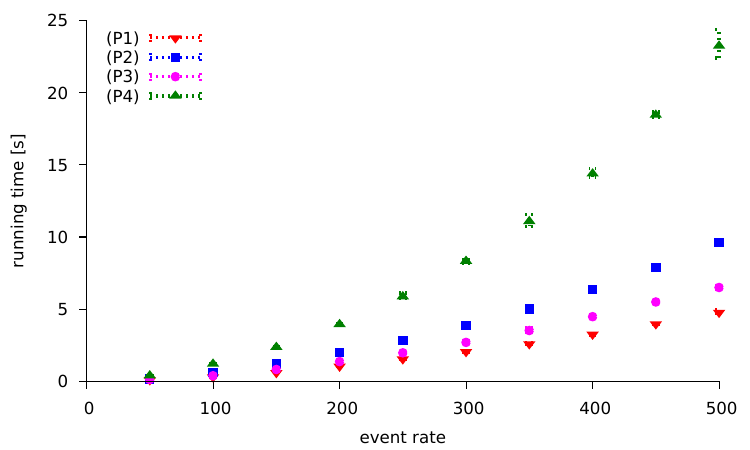}}
  \
  \subfigure[out of order (\MTLfreeze, event rate~$100$)]{\includegraphics[scale=.47]{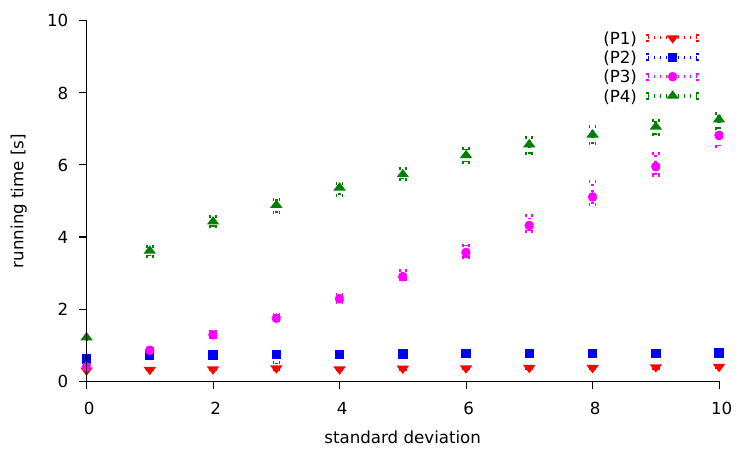}}
  \vspace{-.4cm}
  \caption{Running times over event streams covering a time span of 60
    seconds.  Each data point shows the average over five event
    streams together with the minimum and maximum, which are often
    very close to the average.  Note that for MTL, the event rates are
    10 times higher than for \MTLfreeze.}
  \label{fig:run}
  \vspace{-.4cm}
\end{figure}

The plots in Figure~\ref{fig:run} show the running times for the
formulas in Figure~\ref{fig:formulas}.  Note that the formulas vary in
their temporal and their data ``complexity.''  For MTL, the running
times grow linearly with the event rate and the impact of receiving
events out of order is almost negligible.  This is in contrast to
\MTLfreeze, where running times grow nonlinearly and the impact on
out-of-order events can be significant.  As in the Timescales
experiments memory usage increases for out-of-order events.  However,
the increase is often moderate (usually \POLIMON{} consumes less than
256\,MB of RAM), except for the \MTLfreeze{} out-of-order cases~(P3)
and~(P4) in which the running times also increase significantly.
For putting \POLIMON's performance in relation, we also ran MONPOLY on
corresponding specifications over the ordered event streams.  For MTL,
\POLIMON{} and MONPOLY show similar performance.  For \MTLfreeze,
MONPOLY is in general significantly faster. (P3) is an exception as it
does not have an equivalent counterpart that can be handled by MONPOLY
and even for a variant of it, \mbox{MONPOLY is considerably slower.}

\begin{figure}[t]
  \centering
  \vspace{-.4cm}
  \subfigure[stage running times (Q4)]{\includegraphics[scale=.47]{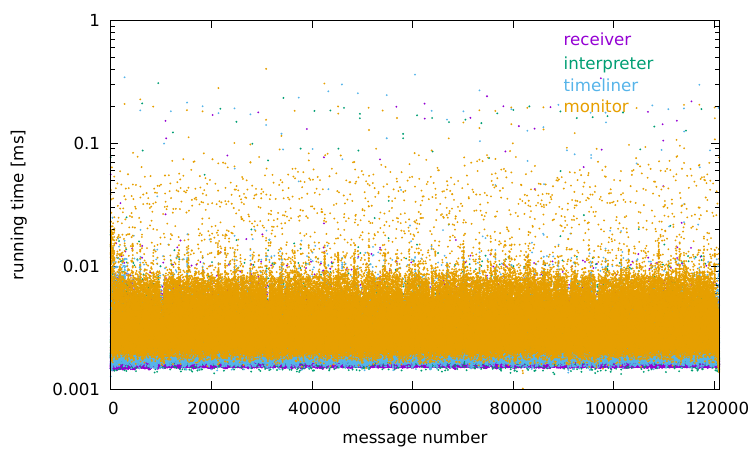}}
  \
  \subfigure[stage running times (P4)]{\includegraphics[scale=.47]{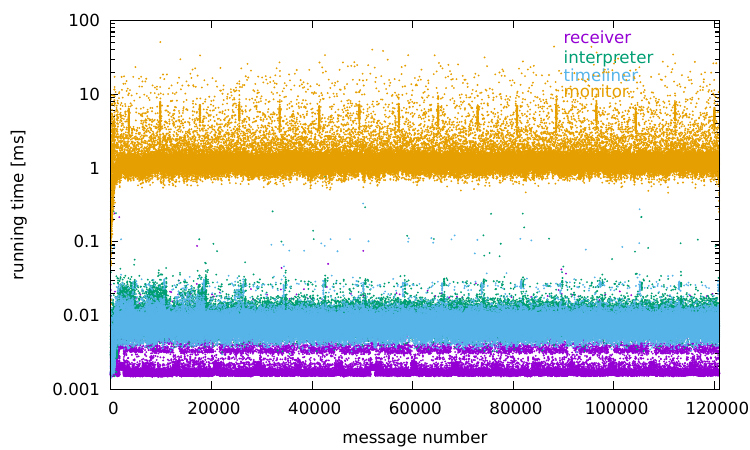}}
  \\[-.2cm]
  \subfigure[stage CDFs (Q4)]{\includegraphics[scale=.47]{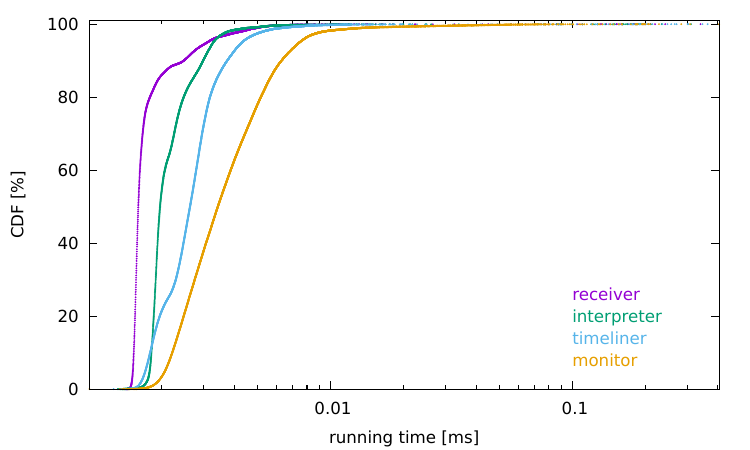}}
  \
  \subfigure[stage CDFs (P4)]{\includegraphics[scale=.47]{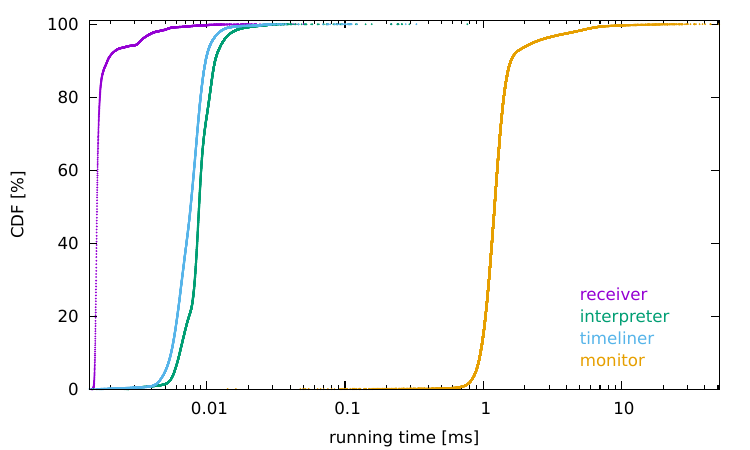}}
  \\[-.2cm]
  \subfigure[graph sizes (Q4)]{\includegraphics[scale=.47]{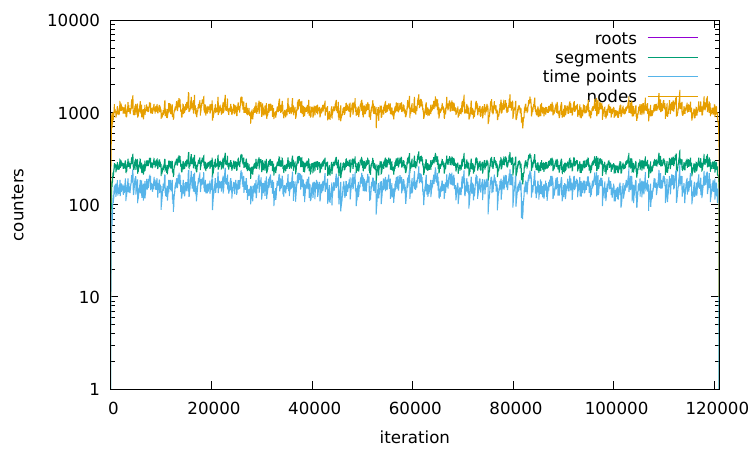}}
  \
  \subfigure[graph sizes (P4)]{\includegraphics[scale=.47]{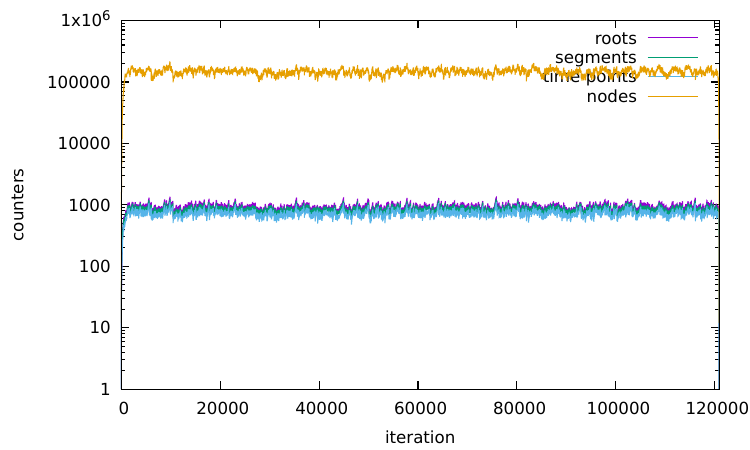}}
  \vspace{-.4cm}
  \caption{Monitor states over an event stream covering a time span of
    10 minutes (event rate~$200$, out-of-order events with standard
    deviation~$1.0$).}
  \label{fig:state}
  \vspace{-.4cm}
\end{figure}

Figure~\ref{fig:state} provides detailed views on individual
\POLIMON{} runs.  The plots on the left-hand side are for the MTL
formula~(Q4) and the ones on the right-hand side are for the
\MTLfreeze{} formula~(P4).  Runs for the other formulas~(P1)--(P3)
and~(Q1)--(Q3) have similar characteristics.

The upper plots ((a) and (b)) show the running times of \POLIMON's
pipeline stages for each event.  For MTL, the stages have similar
running times and each stage is usually executed in a fraction of a
millisecond.  The outliers are mainly caused by Go's garbage
collector.  For \MTLfreeze{}, the monitor stage dominates the running
times.

The middle plots ((c) and (d)) provide another view on the running
times of the pipeline stages. Namely, they show the percentages of a
stage that executed below a time for an event.  Observe that in the
MTL case almost every event is processed under $10\mathit{\mu{}s}$ by
all stages, which is in contrast to the \MTLfreeze{} case, where the
monitor stage takes around $1\mathit{ms}$ for most events, whereas the
other stages take no longer than $10\mathit{\mu{}s}$.  We also remark
that \POLIMON{} in particular benefits in the MTL case from the
multiple CPU cores since the stages run concurrently with a similar
throughput.

The lower plots ((e) and (f)) show the sizes of the data structure.
For \MTLfreeze, the number of nodes in the data structure is
considerable larger than for MTL, which originates from the different
data values.  Observe that although the events are received out of
order, the size of the data structure stays in a certain bound.  This
bound increases when more events are received out of order and thus
more knowledge gaps are closed later.

Overall, \POLIMON{} outperforms its predecessor prototype. It is often
five to ten times faster, sometimes even more,
cf.~\cite{Basin_etal:outoforder_journal}. The implemented algorithms
also scale better, in particular, when events are received out of
order.

\section{Conclusion}
\label{sec:concl}
\vspace{-.2cm}

The \POLIMON{} tool processes out-of-order event streams in real time,
promptly reports verdicts, and provides both soundness and
completeness guarantees.  The underlying monitoring approach and hence
also the tool make the tradeoff of requiring a linear order on system
events by their wall-clock timestamps instead of a partial order on
the events provided by logical clocks.
For MTL specifications, \POLIMON{} handles out-of-order streams
efficiently and its performance is competitive with other monitoring
tools on ordered event streams.  For the setting with data values,
however, the tool's throughput for more complex specifications could
become a bottleneck.  Future work will address this limitation by
further improving the algorithms and their implementation.

\enlargethispage{1cm}
\vspace{-.2cm}
\paragraph{Acknowledgments.}

This work is co-funded by the European Union through the EU H2020
research and innovation programme SPIRS project with Grant Agreement
number~952622.  Views and opinions expressed are however those of the
author(s) only and do not necessarily reflect those of the European
Union.  Neither the European Union nor the granting authority can be
held responsible for them.

\end{document}